\def\versiondate{6 Nov. 2003}
\input math.macros
\input Ref.macros

\checkdefinedreferencetrue
\continuousfigurenumberingtrue
\theoremcountingtrue
\sectionnumberstrue
\forwardreferencetrue
\citationgenerationtrue
\nobracketcittrue
\hyperstrue
\initialeqmacro

\input\jobname.key
\bibsty{myapalike}

\def\cp{\tau}   



\def\cat{{\rm G}}  

\def\changecomma#1,{#1,\,}
\def\bigchangecomma#1,{#1,\;}
\def\leftchangecomma#1,{#1,\ }

\def\T{{\Bbb T}}
\def\frac#1#2{{#1 \over #2}}
\def\cat{{\rm G}}

\def\BLPSusf{\ref b.BLPS:usf/}

\ifproofmode \relax \else\head{{\it J. Phys. A} {\bf 36} (2003), 8361--8365.}
{Version of \versiondate}\fi 
\vglue20pt

\title{High-Precision Entropy Values}
\title{for Spanning Trees in Lattices}
\author{Jessica L. Felker and Russell Lyons}

\abstract{Shrock and Wu have given numerical values for the exponential
growth rate of the number of spanning trees in Euclidean lattices.
We give a new technique for numerical evaluation that gives much more precise
values, together with rigorous bounds on the accuracy.
In particular, the new values resolve one of their questions.
}

\bottomII{Primary 
 05C05, 
60C05. 
Secondary
60K99,  
 05C80, 
82B20. 
}
{Asymptotic complexity, graphs, tree entropy,
uniform spanning forests.}
{Research partially supported by NSF grants DMS-0097813 and DMS-0103897.}

\bsection{Introduction}{s.intro}

Since the time of \ref b.Kirch/, physicists and mathematicians have been
interested in enumerating spanning trees.
One aspect of this endeavor has been to evaluate or to estimate asymptotics of
the growth rate of the number of spanning trees in large graphs.
Additional interest in the asymptotic growth rate arises in ergodic theory,
since the exponential rate is also the entropy of a natural and important
system, the so-called uniform spanning forest. 
See \ref b.Pem:ust/, \ref b.BurPem/, \BLPSusf, \ref
b.Lyons:bird/, and \ref b.Lyons:est/ for explanations and information about
the uniform spanning forest. 
Some modern enumeration efforts include \ref b.BurPem/, \ref b.ShrockWu/, \ref
b.Lyons:est/; see also the references therein.
In many cases, one can express the main term of the asymptotics by an integral
formula.
For example, if $\cp(G)$ denotes the number of spanning trees of a graph $G$
and if $G_n$ are the graphs induced by cubes of side length $n$ in the
hypercubic lattice $\Z^d$, then it is well known (and rederived in both \ref
b.BurPem/ and \ref b.ShrockWu/) that the thermodynamic limit
$$
h_d := \lim_{n \to\infty} {1 \over n^d} \log \cp(G_n)
$$
can be expressed as
$$
h_d=
\int_{\T^d} 
\log\left(2d-2\sum_{i=1}^d \cos2\pi x_i\right)\,dx
=
\log (2d) +
\int_{\T^d} 
\log\left(1 - {1\over d}\sum_{i=1}^d \cos2\pi x_i\right)\,dx
\,.
\label e.ent
$$
It is also well known (due to its connection with the dimer problem)
that $h_2 = 4\cat/\pi$, where $\cat := \sum_{k=0}^\infty
(-1)^k/(2k+1)^2$ is Catalan's constant (see, e.g., 
\ref b.Kas:dim/ or \ref b.Montroll/).
No values of $h_d$ for any $d \ge 3$ are known in simple terms of other known
constants and functions. 
\ref b.ShrockWu/ evaluated these integrals in higher dimensions by numerical
methods and found one particularly intriguing value:
$h_4 = 2.0000(5)$.
They suggested that $h_4$ may be exactly $2$, which
would be quite surprising.
%
%
%
Indeed,
it would be extraordinary for a natural system without parameters to have
a natural-log entropy that is a non-zero integer.
This would have been the first such example to our
knowledge.
However, we shall see that $h_4$ is extraordinarily close to 2, but not, 
in fact, exactly 2. We shall also give more
accurate values for other $h_d$ with rigorous bounds on their accuracy.

The numerical evaluation of $h_d$ is problematic if one wants to use the
formula \ref e.ent/,
due to the difficulty of accurate integration in higher dimensions.
Therefore, \ref b.ShrockWu/ gave an interesting
large-$d$ asymptotic expansion of $h_d$ to order
$1/d^6$; we have given more terms below, to show that not all coefficients are
positive, as one might otherwise believe, and to illustrate that this is indeed
only an asymptotic expansion, not a convergent series:
\begineqalno
h_d =
\log(2d)-\Bigg[
& \frac{1}{4\,d}
+ \frac{3}{16\,d^2}
+ \frac{7}{32\,d^3}
+ \frac{45}{128\,d^4}
+ \frac{269}{384\,d^5}
+ \frac{805}{512\,d^6}
+ \frac{3615}{1024\,d^7}
\cr&+ \frac{23205}{4096\,d^8}
- \frac{144963}{10240\,d^9}
- \frac{2187031}{8192\,d^{10}}
- \frac{40225409}{16384\,d^{11}}
- \frac{1277353077}{65536\,d^{12}}
\cr&- \frac{66817216455}{458752\,d^{13}}
-\frac{271891453119}{262144\,d^{14}}
+O\bigg({1\over d^{15}}\bigg)\Bigg]
\,.
\endeqalno
However,
it is difficult to know how many terms of this expansion to use; \ref
b.ShrockWu/ used this series to
report $h_5 = 2.243$ and $h_6 = 2.437$.
For smaller $d$,
\ref b.ShrockWu/ used numerical integration to
find that $h_3 = 1.6741481(1)$ and that
$h_4 = 2.0000(5)$.
However, the accuracy of $h_3$ is off by several orders of magnitude.

To obtain greater accuracy and enable us to prove that $h_4 \ne 2$, we shall
use a new formula, namely, 
$$
h_d = \log(2d) - \sum_{k = 1}^\infty p_d(k)/k
\,,
\label e.ret
$$
where $p_d(k)$ is the probability that 
simple nearest-neighbor random walk on the hypercubic lattice
$\Z^d$ returns to its starting point after $k$ steps.
(A generalization of this formula is due to \ref b.Lyons:est/.)
Even this formula is slightly problematic to use, since $p_d(k)$ is the sum of
a large number of binomial coefficients for large $k$.
Because of the large number of small terms, it is important to calculate the
sum as an exact rational before converting to a real approximation.
Fortunately, there is a simple recursion formula that enables quicker
computation.
In addition, we explain how to estimate the tail of the series in \ref e.ret/.

In the remainder of this note,
we first state our numerical results and then derive
the simple but crucial
\ref e.ret/.
Next, we explain how to
compute $p_d(k)$ quickly and how to
approximate the error, and finally prove rigorous bounds.
We shall also briefly discuss body-centered cubic lattices.
We end by discussing an alternative approach that was brought to our attention
after a first version of this article was submitted.

\bsection{Results}{s.results}

The numerical
results are, to an accuracy we believe includes all reported digits, in the
following table: 
$$
\hbox{
\vbox{\offinterlineskip
\hrule
\halign{&\vrule#&\strut\quad\hfil#\quad&\vrule#&\strut\quad#\hfil\quad\cr
height2pt&\omit&&\omit&\cr
&$d$\hfil&&\hfil$h_d$&\cr
height2pt&\omit&&\omit&\cr
\noalign{\hrule}
height2pt&\omit&&\omit&\cr
&3 &&1.67338930297&\cr
&4 && 1.999707644517&\cr
&5 && 2.2424880598113&\cr
&6 && 2.43662696200071&\cr
&7 && 2.5986763042&\cr
&8 && 2.73786766385&\cr
&9 && 2.859910142340&\cr
&{10} && 2.968594484443&\cr
&{11} && 3.066571824248&\cr
height2pt&\omit&&\omit&\cr}
\hrule
}
\qquad
\vbox{\offinterlineskip
\hrule
\halign{&\vrule#&\strut\quad\hfil#\quad&\vrule#&\strut\quad#\hfil\quad\cr
height2pt&\omit&&\omit&\cr
&$d$\hfil&&\hfil$h_d$&\cr
height2pt&\omit&&\omit&\cr
\noalign{\hrule}
height2pt&\omit&&\omit&\cr
&{12} && 3.1557714292824&\cr
&{13} && 3.2376421551842&\cr
&{14} && 3.31330031802725&\cr
&{15} && 3.383624540390254&\cr
&{16} && 3.449318935201&\cr
&{17} && 3.510956551787645&\cr
&{18} && 3.56901006528479&\cr
&{19} && 3.62387396384455&\cr
&{20} && 3.67588091671257&\cr
height2pt&\omit&&\omit&\cr}
\hrule
}
}
$$

These arise as follows.
To prove \ref e.ret/,
use the Maclaurin series for $\log(1-z)$ to find
\begineqalno
\int_{\T^d} 
\log\left(1 - {1\over d}\sum_{i=1}^d \cos2\pi x_j\right)\,dx
&=-\sum_{k=1}^\infty {1\over k}
\int_{\T^d}\left[{1\over d}\sum_{i=1}^d\cos2\pi x_j \right]^k\, dx.
\cr&=-\sum_{k=1}^\infty {1\over k}
\int_{\T^d}\left[{1\over 2d}\sum_{i=1}^d \bigg(e^{2\pi i x_j} + e^{-2\pi i x_j}\bigg)
\right]^k\, dx
\cr&=-\sum_{k=1}^\infty {1\over k} p_d(k)
\,.
\endeqalno

Clearly, we have $p_d(k) = 0$ for $k$ odd and $p_1(2k) = {2k \choose
k}/2^{2k}$.
It is well known that $p_2(2k) = p_1(2k)^2$ (e.g., one step of a random walk
in $\Z^2$ can be made by taking one step in each of the directions $\pm
(1/\sqrt2, 1/\sqrt2)$ and $\pm (1/\sqrt2, -1/\sqrt2)$, independently).
For fast computation of other return probabilities, write $f(d, k) :=
(2d)^{2k} p_d(2k)$ for the number of nearest-neighbor walks
of length $2k$ in $\Z^d$ that start
and end at the origin. 
Such a walk has the property that its projection to the first $d_1$
coordinates also starts and ends at the origin, while the number of steps in
the first $d_1$ directions may be any even number between $0$ and $2k$.
Similar reasoning shows that
$$
f(d_1+d_2, k)
=
\sum_{r=0}^k {2k \choose 2r} f(d_1, r) f(d_2, k-r)
\,.
$$
This allows one to reduce the computation of $p_d(\cbuldot)$ to the values of
$p_{\flr{d/2}}(\cbuldot)$ and
$p_{\ceil{d/2}}(\cbuldot)$.

Once we have these values, it is simple to compute
partial sums for \ref e.ret/.
Since all terms of the series are positive, each such partial sum gives a
rigorous upper bound for the true value of $h_d$.
Merely summing the first 13 terms of the series in \ref e.ret/ for $d=4$ gives
a rigorous proof that $h_4 < 2$. 
To get an lower bound for $h_d$, it suffices to bound above the remainder.
It is well known (see, e.g., \ref b.Spitzer/, Section 7) that 
$$
p_d(2k) \sim 2\left({d \over 4 \pi k}\right)^{d/2}
\,.
\label e.std
$$
A more accurate approximation is
$$
p_d(2k) \approx 2\left({d \over 4 \pi k}\right)^{d/2} \bigg(1 - {d \over 8k}
\bigg)
\,,
\label e.better
$$
as shown by \ref b.Ball/.
As this suggests, we believe that the right-hand side of \ref e.std/ is
actually larger than the left-hand side; indeed, this appears to be true for
all $d$ and $k$, not merely for large $k$, though it has been proved only for
large $k$ and small $d$.
That is, we have
$$
p_d(2k) \le 2\left({d \over 4 \pi k}\right)^{d/2}
\label e.bd
$$
for all $k$ when $1 \le d \le 6$ and for all large $k$ (if not all $k$) when
$d \ge 7$; see \ref b.Ball/.
Since the sum over $k \ge r$, any $r > 0$,
of the right-hand sides of either \ref e.std/ or \ref e.better/
can be expressed via the Hurwitz zeta function, for which Euler-Maclaurin
summation approximations are readily available, we obtain the very accurate
values reported in the tables
above by summing relatively few terms. Excellent accuracy is
already available after just 10 terms, but we have used 1000 terms for $3 \le
d \le 6$, 100 terms for $7 \le d \le 10$, and 80 terms for $11 \le d \le 20$.
In addition, by using \ref e.bd/ and partial sums of 1000 terms,
we get the rigorous bounds
\begineqalno
1.6733893024176978\le &h_3 \le 1.6733917596720884\cr
1.9997076445004571 \le &h_4 \le 1.9997076951104138\cr
2.242488059810819 \le &h_5 \le 2.2424880610724065\cr
2.436626962000695 \le &h_6 \le 2.4366269620369234\,.\cr
\endeqalno

One can use similar estimates to improve the accuracy of the asymptotics for
body-centered hypercubic lattices.
As shown by \ref b.ShrockWu/, the exponential growth rate of the number of
spanning trees in $d$ dimensions is 
$$
h^{\rm bcc}_d 
=
d \log 2 - {1\over2} \sum_{k=1}^\infty {1\over k} p_1(k)^d
\,.
\label e.bcc
$$
It is straightforward to show that
$$
p_1(2k) \ge 2\left({1 \over 4 \pi k}\right)^{1/2} \bigg(1 - {1 \over 8k}
\bigg)
\label e.1Dbetter
$$
by use of Stirling's approximation.
We sum 1000 terms of \ref e.bcc/ and bound the remainder.
Using \ref e.bd/ for a lower bound and \ref e.1Dbetter/ for an upper bound, we
find
\begineqalno
1.9901914178466\le &h^{\rm bcc}_3 \le 1.9901914178472\cr
2.732957535468933 \le &h^{\rm bcc}_4 \le 2.7329575354689455\,.\cr
\endeqalno 
These agree with the estimates of \ref b.ShrockWu/, but give about 3 times as
many digits.

After a first version of this article was submitted for publication, Alan
Sokal kindly brought to our attention some related calculations by \ref
b.SokSta/.
Up to a constant, the entropy $h_d$ studied here is equal to the free energy
$g_d(1/d)$ studied there (see equation (A.2) of their paper).
Their formula (A.6) shows, then, that 
$$
h_d = \log (2d) + \int_0^\infty {e^{-t} \over t} \big[1 - I_0(t/d)^d \big] \,dt
\,,
\label e.bessel
$$
where $I_0$ is the modified Bessel function.
In this way, $h_d$ can be estimated by numerical integration in only one
dimension, which can be accomplished very quickly.
The disadvantage, however, is that the integrand decays rather slowly.
As noted in Appendix A.2 of \ref b.SokSta/, one can improve the precision
dramatically by numerical integration up to some cut-off, then symbolic
integration of the tail with an asymptotic formula replacing $I_0$. 
Even so, not all the numerical values reported in \ref b.SokSta/ are correct in all their digits,
as can be seen by comparison with our tables and our rigorous bounds.
The second disadvantage of \ref e.bessel/ is that it is less straightforward
to provide rigorous bounds.
For this purpose, one has to treat carefully the technique of numerical
integration, as well as evaluation and bounding of $I_0$.
Some of this is discussed in Appendix B of \ref b.HaraSlade/.
By comparison, our technique requires only the calculation of rational
numbers, as well as one simple logarithm calculation.

\def\noop#1{\relax}
\input \jobname.bbl

\filbreak
\begingroup
\eightpoint\sc
\parindent=0pt\baselineskip=10pt

Department of Mathematics,
Indiana University,
Bloomington, IN 47405-5701
\emailwww{rdlyons@indiana.edu}
{http://php.indiana.edu/\string~rdlyons/}

and

School of Mathematics,
Georgia Institute of Technology,
Atlanta, GA 30332-0160
\email{rdlyons@math.gatech.edu}

Department of Mathematics, Massachusetts Institute of Technology,
Cambridge, MA  02139
\email{felk@alum.mit.edu}

\endgroup

\bye